\documentclass[english,prb,showpac, manuscript]{revtex4}
\usepackage{mathptmx}

\usepackage{courier}
\usepackage[T1]{fontenc}
\usepackage[latin9]{inputenc}
\usepackage{soul}
\usepackage{xcolor}
\usepackage{amsmath}
\usepackage{graphicx}
\usepackage{amssymb}
\usepackage{esint}

\makeatletter

\providecolor{lyxadded}{rgb}{0,0,1}
\providecolor{lyxdeleted}{rgb}{1,0,0}

\usepackage{cancel}
\usepackage{hyperref}

\usepackage{babel}
\makeatother

\begin{document}

\newcommand{\sgrad}{\boldsymbol{\nabla}_{S}}

\newcommand{\degree}{^{\circ}}

\newcommand{\dif}[1]{\vec{#1}-\vec{#1}'}

\newcommand{\dd}[2]{\delta^{#2}\left(#1-#1'\right)}

\newcommand{\vd}[2]{\delta^{#2}\left(\vec{#1}-\vec{#1'}\right)}

\newcommand{\ds}[1]{\delta\left(#1-#1'\right)}

\newcommand{\grad}{\boldsymbol{\nabla}}

\newcommand{\phik}[3]{e^{#3i#1\cdot#2}}

\newcommand{\intn}[4]{\int_{#1}^{#2}d^{n}\vec{#3}\left(#4\right)}

\newcommand{\height}{\mathcal{H}}

\newcommand{\h}{\height}

\newcommand{\f}{\mathcal{F}}

\newcommand{\e}{\mathcal{E}}

\newcommand{\dx}{\boldsymbol{\Delta}\mathbf{x}}

\newcommand{\av}{\boldsymbol{\alpha}}

\newcommand{\ve}{\boldsymbol{\eta}}

\newcommand{\ez}{\mathcal{E}_{0\degree}}

\newcommand{\sstiff}{{\bf \tilde{\omega}''}}

\newcommand{\ingaas}{\text{In}_{x}\text{Ga}_{1-x}\text{As}/\text{GaAs}}

\newcommand{\gesisi}{\text{Ge}_{x}\text{Si}_{1-x}\text{/Si}}

\newcommand{\fe}[1]{f_{#1}(\vec{k})}

\newcommand{\fee}{\fe{\text{elast.}}}

\newcommand{\fes}{\fe{\text{surf.}}}

\newcommand{\sk}{\sigma(\vec{k})}

\renewcommand{\vec}{\mathbf}

\providecommand{\cs}[1][SOMETHING]{[CITE:#1]\marginpar{C} }

\providecommand{\linkable}{}

\providecommand{\bf}{\boldsymbol}

\preprint{Draft, not for distribution}

\title{Anisotropy and Morphology of Strained III-V Heteroepitaxial Films}

\author{Lawrence H. Friedman}

\affiliation{Penn State University, Department of Engineering Science and Mechanics,
212 Earth and Engineering Sciences Building, University Park, Pennsylvania
16802}

\date{\today}

\begin{abstract}
Strained coherent heteroepitaxy of III-V semiconductor films such
as $\text{In}_{x}\text{Ga}_{1-x}\text{As}/\text{GaAs}$ has potential
for electronic and optoelectronic applications such as high density
logic, quantum computing architectures, laser diodes, and other optoelectronic
devices. Crystal symmetry can have a large effect on the morphology
of these films and their spatial order. Often the formation of group
IV strained heterostructures such as Ge deposited on Si is analyzed
using analytic models based on the Asaro-Tiller-Grinfeld instability.
However, the governing dynamics of III-V 3D heterostructure formation
has different symmetry and is more anisotropic. The additional anisotropy
appears in both the surface energy and the diffusivity. Here, the
resulting anisotropic governing dynamics are studied to linear order.
The resulting possible film morphologies are compared with experimentally
observed $\text{In}_{x}\text{Ga}_{1-x}\text{As}/\text{GaAs}$ films.
Notably it is found that surface-energy anisotropy plays a role at
least as important as surface diffusion anisotropy if not more so,
in contrast to previous suppositions.
\end{abstract}


\maketitle
Strained coherent heteroepitaxy of $\ingaas$ at high temperatures
can lead to dense somewhat correlated 3D film growth, for example,
rolls (Figs.~2c and~d in Ref.~\citep{Chokshi:2000dc}), as well
as correlated dense arrays of self-assembled quantum dots (Fig.~1
in Ref.~\citep{Liang:2006fk}) and quantum dot chains in multilayers
(Fig.~1a in Ref.~\citep{Schmidbauer:2006qe}). These structures
are of great interest for the next generation of electronic and optoelectronic
materials due to their quantum confinement effects,~\citep{Bimberg99}
and they are an excellent example of nanoscale self-assembly that
can augment or replace traditional lithographic techniques. The density
and degree of correlation in these film morphologies suggest that
their growth is barrierless and cooperative. Furthermore, these structures
form at high temperature, suggesting that the surfaces might be thermally
roughened. Thus, the film morphology is likely governed by the Asaro-Tiller-Grinfeld
(ATG) instability or related mechanism; whereby surface diffusion
is driven by changes in elastic and surface energy.~\citep{Asaro:1972qq,Grinfeld:1986tg,Srolovitz:1989fu,Spencer:1993vt,Golovin:2003ms}
Unlike modeling group IV structures (\emph{e.g.} $\gesisi$),~\citep{Ozkan:1999gf,Obayashi:1998fk}
modeling the growth of III-V heterostructures as an ATG-like process
requires full consideration of three contributions to anisotropic
pattern formation, namely elastic anisotropy, surface-energy density
anisotropy and diffusion anisotropy. The role of anisotropic effects
has been increasingly recognized in nanoscale epitaxial self-assemly.~\citep{Pradhan:2004uo,Zandvliet:2007sp,Thayer:2004bs}
Here, a linear model is presented that has appropriate symmetries
for III-V structures generally, and $\ingaas$ in particular. 

Similar models have been largely applied to group IV heteroepitaxy,
such as $\gesisi$, that have four-fold rotational symmetry of the
crystal surface, but III-V systems have only two-fold rotational symmetry.
The experimental observations bear out this difference. $\ingaas$
structures grown on $(001)$ surfaces form rows or rolls that are
aligned closer to the $[1\bar{1}0]$ direction than the $[110]$ direction.
On the other hand, $\gesisi$ structures form a four-fold symmetric
pattern with alignment along the $\left\langle 100\right\rangle $
directions.~\citep{Ozkan:1997uq} This spatial alignment of $\ingaas$
structures closer to the $[1\bar{1}0]$ direction has been attributed
to differences in surface diffusivity~\citep{Schmidbauer:2006qe,Liang:2006fk,Chokshi:2000dc}
and possibly to packing effects and surface faceting.~\citep{Chokshi:2000dc}
To test these hypotheses, one must obtain similar structures from
a complete model of surface evolution. A linear analysis of surface
diffusion based on the original ATG instability has been used to explain
anisotropic patterning in $\gesisi$ heteroepitaxy,~\citep{Ozkan:1999gf,Obayashi:1998fk},
and it would be remiss not to apply it to $\ingaas$. In group IV
systems, the symmetry requirements forbid anisotropic surface-energy
and diffusion effects to linear order.~\citep{Friedman:2007kx} Thus,
for $\gesisi$ the $\left\langle 100\right\rangle $ alignment comes
from elastic anisotropy. The lower symmetry of III-V (001) surfaces
allows surface-energy and diffusion anisotropy effects; thus, all
three contributions interact to determine the alignments of 3D morphological
features. 

In the following, a continuum linear model of anisotropic film evolution
is presented including the general framework, then energetic and diffusional
sources of anisotropy. Then the consequences for film morphology are
presented and compared with experimental observations.

Early stage small fluctuations in film height are an important determiner
for final film morphology~\citep{Golovin:2003ms,Spencer:1993vt,Srolovitz:1989fu}
and can be analyzed using linear dynamics of the film height that
is a sum of the average film height ($\bar{\h}$) and film height
fluctuations ($h(\vec{x},t)$), $\h(\vec{x},t)=\bar{\h}+h(\vec{x},t)$.
The normal modes of the film evolution are periodic; thus, Fourier
components of the height fluctuations are used; $h_{\vec{k}}=A^{-1}\int d^{2}\vec{x}\, e^{-i\vec{k}\cdot\vec{x}}h(\vec{x})$
and $h\left(\vec{x}\right)=\sum_{\vec{k}}e^{i\vec{k}\cdot\vec{x}}h_{\vec{k}}$,
where $A$ is the area of the substrate, and periodic boundary conditions
are assumed.  Following Ref.~\citep{Friedman:2007if}, each Fourier
component is initially governed by energy dissipation and thermal
fluctuations,

\begin{eqnarray}
\partial_{t}h_{\vec{k}} & = & \sk h_{\vec{k}}+i\vec{k}\cdot\sqrt{2k_{b}T\mathbf{D}}\cdot\eta_{\vec{k}},\nonumber \\
\sk & = & -\left(\vec{k}\cdot\mathbf{D}\cdot\vec{k}\right)f(\vec{k}),\label{eq:dispersion}\end{eqnarray}
where $\sk$ is the dispersion relation, $\mathbf{D}$ is the surface
diffusivity, $\eta_{\vec{k}}(t)$ is a normally distributed stochastic
variable with mean $\left\langle {\bf \eta}_{\vec{k}}(t)\right\rangle =0$,
and variance $\left\langle {\bf \eta}_{\vec{k}}(t){\bf \eta}_{\vec{k}'}^{*}(t')\right\rangle =A^{-1}\delta_{\vec{k}\vec{k}'}\ds{t}$.
$\delta_{\vec{k}\vec{k}'}$ is the Kronecker delta, $\delta(t-t')$
is the Dirac delta, and $\fe{}$ is the energy cost function, formally
defined as the second derivative of the free energy per unit area
($\f/A$) with respect to each Fourier component,  \begin{equation}
\fe{}=\left.\partial_{h_{\vec{k}}}\partial_{h_{\vec{k}}^{*}}(\f/A)\right|_{h_{\vec{k}}=0}.\label{eq:cost_function}\end{equation}
 From $\sk$, one can determine various length and time scales as
well as pattern orientation and alignment.

The film free energy cost function consists of two terms due to elastic
strain and surface energy $\fe{}=\fe{\text{elast.}}+\fe{\text{surf. }}$.
For now, the wetting energy contibution~\citep{Beck:2004yq} is neglected
for simplicity. For an (001) surface, elastic anisotropy leads to
4-fold-symmetric surface energetics and dynamics, and surface energy
anisotropy leads to 2-fold symmetric energetics and dynamics as explained
below.

The effect of elastic anisotropy alone has been discussed previously,~\citep{Ozkan:1999gf,Obayashi:1998fk,Friedman:2007fk,Friedman:2007kx}
and the most important results are summarized below. For (001) surfaces,
the elastic part of the energy cost function can be approximated as~\citep{Friedman:2007kx}\begin{equation}
\fee=-\left[\e_{0\degree}\cos^{2}\left(2\theta_{\vec{k}}\right)+\e_{45\degree}\sin^{2}\left(2\theta_{\vec{k}}\right)\right]k,\label{eq:f-elast}\end{equation}
where $\ez$ and $\e_{45\degree}$ are constants, $\theta_{\vec{k}}$
is the angle between $\vec{k}$ and the $[100]$ direction, and $k=\left\Vert \vec{k}\right\Vert $.
$f_{\text{elast.}}$ is most negative along the $\left\langle 100\right\rangle $
directions; thus ripples perpendicular to these directions release
the most elastic energy. Without competing anisotropic effects, the
anisotropy of $f_{\text{elast.}}$ causes initial alignment of 3D
gratings along the $\left\langle 100\right\rangle $ directions.~\citep{Ozkan:1999gf,Obayashi:1998fk}
For InAs, $\e_{0\degree}=8.13\times10^{9}\text{ erg/cm}^{3}$, and
$\e_{45\degree}=6.92\times10^{9}\text{ erg/cm}^{3}$.

One can derive the general form of $f_{\text{surf.}}$ as follows.
Assume that the total free energy is an integral in the $\vec{x}-\text{plane}$
over a local free energy kernel $\omega$ that depends on the surface
orientation $\grad\h$, $\f_{\text{surf.}}=\int d^{2}\vec{x}\,\omega\left(\grad\h\left(\vec{x}\right)\right).$
In terms of the surface energy density $\gamma$, $\omega=\gamma(\grad\h)[1+\left(\grad\h\left(\vec{x}\right)\right)^{2}]^{1/2}$.
Using Eq.~\ref{eq:cost_function}, the free energy cost function
is \begin{equation}
\fes=\vec{k}\cdot{\bf \tilde{\omega}''}\cdot\vec{k},\label{eq:f-surf-wet}\end{equation}
where ${\bf \tilde{\omega}''}=\partial_{\grad\h}^{2}\omega|_{\grad\h=\vec{0}}=\gamma(\vec{0}){\bf \tilde{I}}+{\bf \tilde{\gamma}''}(\vec{0})$.
${\bf \tilde{\omega}''}$ acts like a surface stiffness stabilizing
short wavelengths.~\citep{Friedman:2007kx}

From Eq.~\ref{eq:f-surf-wet}, one can determine the possible form
of any anisotropy. $\sstiff$ is a rank 2, dimension 2, symmetric
tensor, and it has two eigenvalues and eigenvectors. Due to crystal
symmetry, namely reflections through the $(110)$ and $\left(1\bar{1}0\right)$
planes, the eigenvectors of $\sstiff$ must be in the $[110]$ and
$[1\bar{1}0]$ directions with eigenvalues, $\omega_{[110]}$ and
$\omega_{[1\bar{1}0]}$ . The surface energy anisotropy coefficient
can be defined as $\delta_{\text{surf.}}=\omega_{[1\bar{1}0]}/\omega_{[110]}$
so that $\sstiff=\omega_{0}\left(\mathbf{n}_{[110]}\mathbf{n}_{[110]}+\delta_{\text{surf.}}\mathbf{n}_{[1\bar{1}0]}\mathbf{n}_{[1\bar{1}0]}\right)$,
where $\omega_{0}=\omega_{[110]}$, $\mathbf{n}_{[\dots]}$ is the
unit vector in the specified direction, and the vector products are
outer products.  Thus, $\delta_{\text{surf.}}=1$ indicates an isotropic
surface stiffness, and $\delta_{\text{surf.}}>1\,(<1)$ indicates
an enhanced stiffness in the $\pm[1\bar{1}0]\,(\pm[110])$ directions.
In terms of the surface energy density, $\delta_{\text{surf.}}=(\gamma(\vec{0})+\gamma_{[1\bar{1}0]})/(\gamma(\vec{0})+\gamma_{[110]}),$
where $\gamma_{[\dots]}$ are the eigenvalues of ${\bf \tilde{\gamma}''}$.

\begin{figure*}
\begin{centering}
\includegraphics{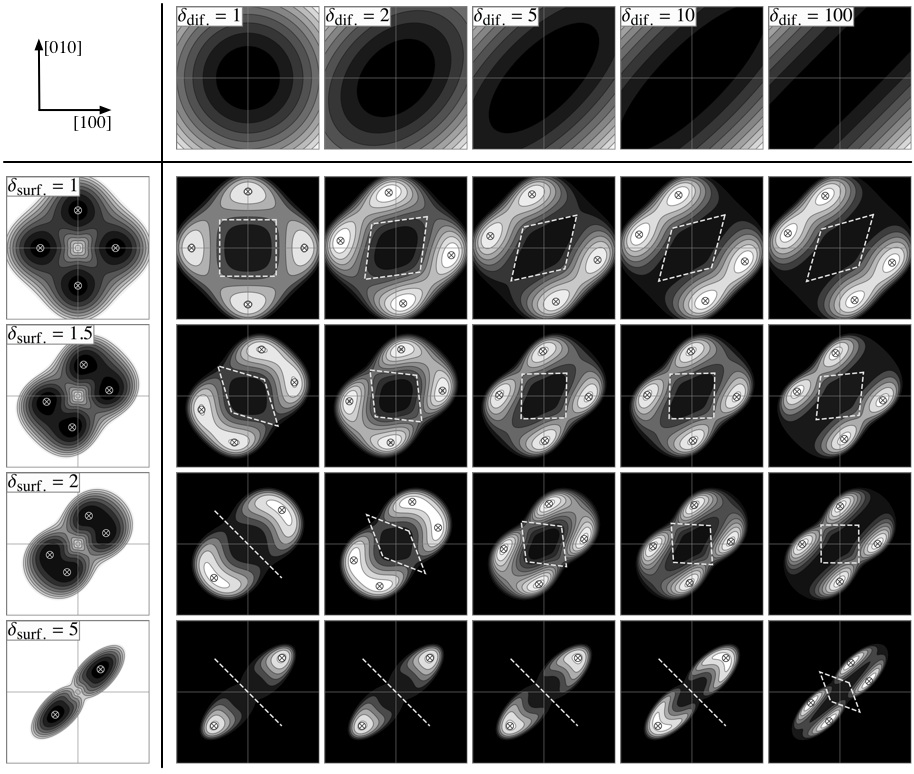}
\par\end{centering}

\caption{\label{fig:Muliplication-Table}\emph{Graphical {}``multiplication
table''} showing effects of diffusional and surface energy anisotropy.
Top row shows $\left(\vec{k}\cdot\mathbf{D}\cdot\vec{k}\right)/D_{0}$
for various surface-diffusion anisotropies $\delta_{\text{dif.}}$.
Left column shows $\fe{}$ for various surface-stiffness anisotropies
$\delta_{\text{surf.}}$. Interior grid shows resulting dispersion
relation $\sk$ (Eq.~\ref{eq:dispersion}). Horizontal direction
($\rightarrow$) is $[100]$. Vertical direction ($\uparrow$) is
$[010]$. All axes span $\pm1\,\text{rad/nm}$. $\otimes$'s indicate
$\vec{k}_{E}$ for $\fe{}$ and $\vec{k}_{0}$ for $\sk$ plots. Dashed
lines perpedicular to $\vec{k}_{0}$ indicate mean alignment of grating
rows or rolls (Fig.~\ref{fig:Simulated-film-heights}).}

\end{figure*}

The exact values of $\omega_{0}$ and $\delta_{\text{surf.}}$ are
not known, so $\omega_{0}$ will be taken to be about $770\text{ erg/cm}^{2}$.~\citep{Pehlke:1996kx}
Then, taking in turn $\delta_{\text{surf.}}=1$, $1.5$, $2$, and
$5$ the total energy cost functions are shown on the left-hand column
of Fig.~\ref{fig:Muliplication-Table}. Only values of $\delta_{\text{surf.}}\geq1$
are shown, corresponding to enhanced stiffness in the $\pm[1\bar{1}0]$
directions. 

In all cases, the energy cost function has negative minima indicating
the wavevectors of the energetically most unstable modes, $\vec{k}_{E}$.
In agreement with previous findings,~\citep{Ozkan:1999gf,Obayashi:1998fk}
for the isotropic surface energy case, $\vec{k}_{E}$ point along
the $\left\langle 100\right\rangle $ directions with magnitude $k_{E}=\mathcal{E}_{0\degree}/\left(2\omega_{0}\right)$
. However, for $\delta_{\text{surf.}}>1$, the peak wave vectors move
towards the $\pm[110]$ directions, and the magnitudes decrease so
that $k_{E}<\mathcal{E}_{0\degree}/\left(2\omega_{0}\right)$. For
$\delta_{\text{surf.}}>2.4$, the peaks actually merge to form one
peak with $\vec{k}_{E}$ along the $\pm[110]$. An opposite trend
would be found for $\delta_{\text{surf.}}<1$ (not shown).

\begin{figure*}
\noindent \begin{centering}
\includegraphics{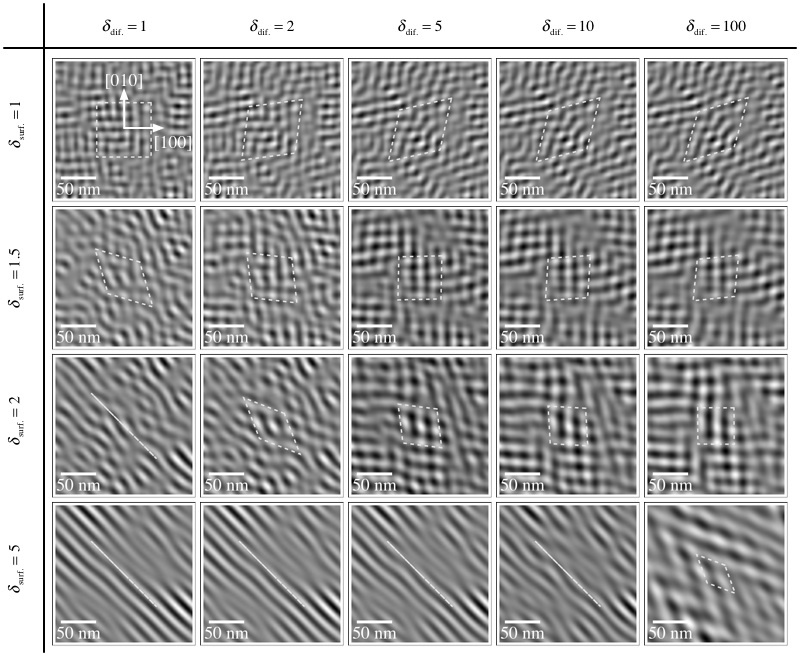}
\par\end{centering}

\caption{\label{fig:Simulated-film-heights}{}``Simulated'' film heights
for various values of $\delta_{\text{dif.}}$ and $\delta_{\text{surf.}}$
corresponding to $\sk$'s appearing in Fig.~\ref{fig:Muliplication-Table}.
All figures oriented with $[100]$ horizontal ($\rightarrow$) and
$[010]$ vertical ($\uparrow$). Dashed lines show mean grating or
roll alignment predicted by $\sk$.}

\end{figure*}

In addition to energetics, one must consider dynamics. The surface
diffusivity is a rank 2, dimension 2 tensor, and it must have the
same symmetry as the crystal surface. Thus, it has two eigenvectors
and eigenvalues corresponding to the $[110]$ and $[1\bar{1}0]$ directions
just as ${\bf \tilde{\omega}}''$ does. Let $\delta_{\text{dif.}}=D_{[110]}/D_{[1\bar{1}0]}$and
$D_{0}=D_{[110]}$. Then, $\mathbf{\tilde{D}}=D_{0}(\delta_{\text{dif.}}^{-1}\mathbf{n}_{[110]}\mathbf{n}_{[110]}+\mathbf{n}_{[1\bar{1}0]}\mathbf{n}_{[1\bar{1}0]})$.
$\delta_{\text{dif.}}>1$ indicates slower diffusion in the $[110]$
direction, which is believed for As terminated $(001)B$ InAs.~\citep{Shiraishi:1992uq}
The top row of Fig.~\ref{fig:Muliplication-Table} shows $\left(\vec{k}\cdot\mathbf{\tilde{D}}\cdot\vec{k}\right)/D_{0}$
for $\delta_{\text{dif.}}=1$, $2$, $5$, $10$ and $100$.

Three contributions to anisotropy in surface morphology have been
discussed. The first two, elastic anisotropy and surface-stiffness
anisotropy contribute to an anisotropic energy cost function, $\fe{}$.
The last, diffusional anisotropy, multiplies $\fe{}$ to give an anisotropic
dispersion relation $\sk$ (Eq.~\ref{eq:dispersion}). The peaks
in $\sk$ give the fastest growing perturbation wavevectors, $\vec{k}_{0}$.
Fig.~\ref{fig:Muliplication-Table} shows a figurative multiplication
table of the combined effects of surface-stiffness anisotropy and
diffusive anisotropy on the dispersion relation. The left column and
top row show the surface-stiffness and diffusivity anisotropy respectively
while the inner grid shows the combined effects. Resulting real-space
morphological alignements for each case will perpendicular to $\vec{k}_{0}$
and are shown by dashed lines in Fig.~\ref{fig:Muliplication-Table}.

From examination of Fig.~\ref{fig:Muliplication-Table} and calculations
using other $\delta_{\text{surf.}}$ and $\delta_{\text{dif.}}$ values
(not shown), one finds:

\begin{enumerate}
\item When only elasticity is anisotropic ($\delta_{\text{surf.}}=\delta_{\text{dif.}}=1)$,
the peaks are oriented along the $\left\langle 100\right\rangle $
directions, the same as the peaks in $\fe{}$, but the magnitudes
$k_{0}=\left|\vec{k}_{0}\right|=\left(3/2\right)k_{E}=3\omega_{0}/\left(4\mathcal{E}_{0\degree}\right)$
as expected.~\citep{Golovin:2003ms,Friedman:2007kx}
\item When $\delta_{\text{dif.}}>1$ and $\delta_{\text{surf.}}=1$, diffusivity
is greater in the {}``fast'' $[1\bar{1}0]$ direction, but with
isotropic surface stiffness the peaks move towards the fast $\pm[1\bar{1}0]$
directions, and the magnitude $k_{0}$ decrease.
\item When $\delta_{\text{dif.}}=1$, but $\delta_{\text{surf.}}$ $>1$,
the peaks move towards the {}``slow'' $\pm[110]$ directions. In
particular, for $\delta_{\text{surf.}}>1.94$, the pairs of peaks
in the $\pm[110]$ directions each merge into one peak exactly along
the $\pm[110]$ directions. Note that the energetic minima at $\vec{k}_{E}$
merge at values, $\delta_{\text{surf.}}>2.4$.
\item When $\delta_{\text{dif.}}>1$, and $\delta_{\text{surf.}}>1$, the
two effects compete. For example, for $\delta_{\text{dif.}}=100$
and $\delta_{\text{surf.}}=2$, the peak positions $\vec{k}_{0}$
appear almost exactly along the the $\left\langle 100\right\rangle $
axes, and for large $\delta_{\text{surf.}}$ and $\delta_{\text{dif.}}$,
for example, $\delta_{\text{surf.}}=5$ and $\delta_{\text{dif.}}=100$,
the merged peaks in $\fe{}$ are split by the anisotropic surface
diffusivity back into four peaks that appear near the slow $\pm[110]$
directions.
\end{enumerate}

The initial film morphology will be quasiperiodic with reciprocal
lattice vectors given by the peaks in $\sk$ (Fig.~\ref{fig:Muliplication-Table}).
For visualization, one can sample film height fluctuations from the
appropriate statistical distribution. Following Ref.~\citep{Friedman:2007if},
Eq.~\ref{eq:dispersion} is used to find the ensemble means and variances
of the individual film height Fourier components. Taking the ensemble
average of Eq.~\ref{eq:dispersion}, $\partial_{t}\left\langle h_{\vec{k}}\right\rangle =\sk\left\langle h_{\vec{k}}\right\rangle $.
Since initially $\left\langle h_{\vec{k}}\right\rangle =0$, $\langle h_{\vec{k}}\rangle=0$
for all $t$. Using the stochastic chain rule for $\partial_{t}\left|h_{k}\right|^{2}$
and taking an ensemble average, $\partial_{t}\left\langle \left|h_{k}\right|^{2}\right\rangle =2\sigma_{\vec{k}}\left\langle \left|h_{k}\right|^{2}\right\rangle +2k_{b}T\left(\vec{k}\cdot\mathbf{D}\cdot\vec{k}\right)/A$.
Solving, $\langle|h_{\vec{k}}|^{2}\rangle=k_{b}T\left(e^{2\sk t_{\text{large}}}-1\right)/[A\, f(\vec{k})]$,
where $t_{\text{large}}$ is chosen so that the r.m.s film height
fluctuation is an appropriate value for small fluctuations, $h_{\text{r.m.s}}=\left(\sum_{\vec{k}}\langle|h_{\vec{k}}|^{2}\rangle\right)^{1/2}=2.5\text{ \AA}$.

Fig.~\ref{fig:Simulated-film-heights} shows sampled initial film
morphologies corresponding to the dispersion relations $\sk$ from
Fig.~\ref{fig:Muliplication-Table}. Dashed lines in both figures
are perpendicular to $\vec{k}_{0}$ and show the mean orientations
of either grating rows or rolls. Dispersion relations ($\sk$) with
four peaks lead to quasiperiodic gratings that are suitable precursors
to self-assembled quantum dots. Dispersion relations with two peaks
lead to roll structures. These structures, however, might evolve significantly
during later development perhaps in a fashion similar to ripening.
Such evolution might depend in detail on the surface energy density
$\gamma(\grad\h)$ to higher than linear order and is thus a subject
for future investigation.

Figs.~\ref{fig:Muliplication-Table} and~\ref{fig:Simulated-film-heights}
indicate that fast diffusion in the $[1\bar{1}0]$ direction alone
can not be responsible for the observation of film morphologies aligned
either along or close to the $[1\bar{1}0]$ direction.~\citep{Chokshi:2000dc,Liang:2006fk,Schmidbauer:2006qe}
In fact, it gives the opposite result, alignment close to the slow
$[110]$ direction. $[1\bar{1}0]$-aligned patterns occur either for
moderate values of $\delta_{\text{surf.}}$ and small or zero $\delta_{\text{dif.}}$,
or for the case where both $\delta_{\text{surf.}}$ and $\delta_{\text{dif.}}$
are large. It is generally accepted that the $[1\bar{1}0]$ direction
is the fast diffusion direction~\citep{Shiraishi:1992uq} suggesting
that the latter case is more likely than the former. As mentioned
above, $\delta_{\text{surf.}}<1$ would have the opposite effect;
thus observations of SAQD arrangements can be used to infer that $\delta_{\text{surf.}}>1$.
A Fourier transform of Fig.~1 from Ref.~\citep{Liang:2006fk} has
broad peaks that overlap the slow $[110]$ directions significantly.
Perhaps the spectral support along the $[110]$ direction is initially
suppressed by diffusional anisotropy and appears later due to non-linear
effects which will be the subject of future investigations. The other
mentioned images from Refs.~\citep{Chokshi:2000dc,Schmidbauer:2006qe}
have spectral support that is close to the $[110]$ direction but
with less overlap. 

A linear surface evolution model was presented that incorporates elastic,
surface energy and diffusional anisotropy. It was found that fast
diffusion in the $[1\bar{1}0]$ direction gives initial morhpology
aligned along the slow $[110]$ direction. Thus, oberved $[1\bar{1}0]$
directed alinement of morphology~\citep{Chokshi:2000dc,Schmidbauer:2006qe,Liang:2006fk}
must be caused in whole or in part by surface energy anisotropy. Future
non-linear modeling is needed to confirm this observation and may
reveal even more complicated mechanisms.

\begin{acknowledgments}
Thanks to G. Salamo and Zh. Wang (U. of Arkansas) and J. Millunchick
(U. of Michigan) for useful discussion.
\end{acknowledgments}
\bibliographystyle{apsrev}
\bibliography{Bib-Friedman}

\end{document}